\documentclass[12pt,a4paper]{article}
\usepackage{amsmath,amssymb}
\newcommand{\ba}{\begin{array}}
\newcommand{\ea}{\end{array}}
\newcommand{\pa}{\partial}
\newcommand{\la}{\lambda}
\newcommand{\ep}{\epsilon}
\newcommand{\no}{\nonumber}
\newcommand{\tp}{\tilde{p}}
\newcommand{\tq}{\tilde{q}}
\newcommand{\tD}{\tilde{D}}
\newcommand{\tS}{\tilde{S}}
\newcommand{\bp}{\bar{p}}
\newcommand{\bz}{\bar{z}}
\newcommand{\bt}{\bar{t}}
\newcommand{\vphi}{\varphi}
\newcommand{\mF}{\mathcal{F}}

\begin{document}

\title{Solutions for real dispersionless Veselov-Novikov hierarchy}
\author{Jen-Hsu Chang and Yu-Tung Chen \\ \\
Department of Computer Science, \\
National Defense University, Taoyuan, Taiwan \\ \\
E-mail: jhchang@ndu.edu.tw}

\maketitle

\begin{abstract}
We investigate  the dispersionless Veselov-Novikov (dVN) equation
based on the framework of dispersionless two-component BKP
hierarchy. Symmetry constraints for real dVN system are
considered. It is shown that under symmetry reductions, the
conserved densities are therefore related to the associated Faber
polynomials and can be solved recursively. Moreover, the method of
hodograph transformation as well as the expressions of Faber
polynomials are used to find exact real solutions of the  dVN
hierarchy.
\end{abstract}

\maketitle
%
\section{Introduction}
%
The Veselov-Novikov equation
\begin{equation}
 u_{\tau }= (uV)_z + (u\bar{V})_{\bz} + u_{zzz} + u_{\bz\bz\bz}, \quad V_{\bz}=-3u_z
\label{VNeq}
\end{equation}
was invented in \cite{VN84} as a certain two-dimensional integrable extension of the KdV equation.
Here $z=x+iy$ and the subscripts $z,\bz,\tau$ denote partial derivatives.
The important subclass of this equation is the so-called dispersionless Veselov-Novikov (dVN) equation
that has been considered \cite{Kri88,KM02,BKM06} by taking the quasi-classical limit of (\ref{VNeq}),
in which the dispersion effect had been dropped. Namely,
\begin{equation}
 u_{\tau}=(uV)_z+(u\bar{V})_{\bz}, \quad V_{\bar{z}}=-3u_z.
\label{dVNeq}
\end{equation}
Recently, it was demonstrated that in \cite{KM02,KM04a,KM04b,KM05} the dVN hierarchy
is amenable to the semiclassical $\bar{\pa}$-dressing method.
Also, the dVN equation and dVN hierarchy have appeared in aspects of
symmetries and relevant in the description of geometrical optics phenomena
\cite{BKM06,KM04a,KM05,KM04b}.
In \cite{BKM06}, some symmetry constraints for dVN equations were
proposed to be efficient ways of construction of reductions (see
also \cite{BK04,BK05,Ch07} for symmetry constraints of
dispersionless integrable equations). It was also shown that the
dVN equation can be reduced into (1+1)-dimensional hydrodynamic
type systems under the symmetry constraint.

The dispersionless Hirota equations for the two-component BKP system was first derived
by Takasaki \cite{Tak06} as the dispersionless limit of the differential Fay identity.
Later, the Hirota equations was rederived \cite{CT06} from the method
of kernel formulas provided by Carroll and Kodama \cite{CK95}.
As observed in \cite{KM04a,KM04b,Tak06}, the Hamilton-Jacobi equation arising from extra equation
of these Hirota equations can be related to the dVN equation as well as the
Eikonal equation in the geometrical optics limit of Maxwell equations.
Inspire by these observations, we are interested in connecting dVN hierarchy to
the dispersionless two-component BKP (2-dBKP) hierarchy.
In this paper, we study the dVN equation based on the framework of
the 2-dBKP hierarchy \cite{Tak06} (see also \cite{CT06} for the
extended dBKP hierarchy). The Hirota equations of 2-dBKP provide
an effective way of constructing Faber polynomials
\cite{Pom75,Teo03} and the associated Hamilton-Jacobi equations.
Real symmetry constraints will be imposed to find the real
solutions of the dVN hierarchy.

This paper is organized as follows.
In section \ref{Sec-dVN}, we recall the 2-dBKP hierarchy and give
an identification of hierarchy flows for the 2-dBKP system
and the dVN hierarchy.
In section \ref{Sec-Faber}, we derive Faber polynomials
of the dVN hierarchy by means of its Hirota equations.
In particular, we obtain the recursion formulas of the Faber polynomials.
In section \ref{Sec-sym}, under symmetry constraint for the dVN hierarchy,
we show that the corresponding Faber polynomials characterize
the second derivatives of free energy and the conserved densities.
In section \ref{Sec-hodo}, we present the hodograph solutions
for the dVN hierarchy by choosing some suitable initial data.
The solutions for $S$ function are also given as examples.
In section \ref{Sec-2N}, we discuss $2N$-component symmetry
constraint for the dVN hierarchy and derive the corresponding
conserved densities.
Section \ref{Sec-Con} is devoted to the concluding remarks.

%
\section{The dVN hierarchy}\label{Sec-dVN}
%
The 2-dBKP system can be characterized by the following Hirota equations \cite{Tak06, CT06}
\begin{eqnarray}
 \frac{p(\la)-p(\mu)}{p(\la)+p(\mu)} &=& \exp(-D(\la)S(\mu)),
\label{EdBKP1} \\
 \frac{\tp(\la)-\tp(\mu)}{\tp(\la)+\tp(\mu)} &=& \exp(-\tD(\la)\tS(\mu)),
\label{EdBKP2} \\
 \frac{p(\la)-\tq(\mu)}{p(\la)+\tq(\mu)} &=& \exp(-D(\la)\tS(\mu))=\exp (D(\la)\tD(\mu)\mF),
\label{EdBKP3} \\
 \frac{\tp(\mu)-q(\la)}{\tp(\mu)+q(\la)} &=& \exp(-\tD(\mu)S(\la))=\exp (\tD(\mu)D(\la)\mF),
\label{EdBKP4}
\end{eqnarray}
where the generating functions $S(\la), \tS(\la)$ are defined by
\[
S(\la) = \sum_{n=0}^{\infty}t_{2n+1}\la^{2n+1}-D(\la)\mF, \quad
\tS(\la) = \sum_{n=0}^{\infty}\tilde{t}_{2n+1}\la^{2n+1}-\tD(\la)\mF,
\]
and
$D(\la)=\sum_{n=0}^{\infty}\frac{2\la^{-2n-1}}{2n+1}\pa_{t_{2n+1}}$,
$\tD(\la)=\sum_{n=0}^{\infty}\frac{2\la^{-2n-1}}{2n+1}\pa_{\tilde{t}_{2n+1}}$
denote the  vertex operators \cite{BK05}; morever, $p(\la),
q(\la), \tilde{p}(\la), \tilde{q}(\la)$ are defined by
\begin{eqnarray}
&& p(\la) = \frac{\pa S(\la)}{\pa t_1} = \la - D(\la)\pa_{t_1}\mF, \quad
  q(\la) = \frac{\pa S(\la)}{\pa \tilde{t}_1} = - D(\la)\pa_{\tilde{t}_1}\mF, \\
&& \tp(\la) = \frac{\pa \tS(\la)}{\pa \tilde{t}_1} = \la - \tD(\la)\pa_{\tilde{t}_1}\mF, \quad
  \tq(\la) = \frac{\pa \tS(\la)}{\pa t_1} = - \tD(\la)\pa_{t_1}\mF.
\end{eqnarray}
By equating Eqs. (\ref{EdBKP3}) and (\ref{EdBKP4}), one has the
equation \cite{Tak06}: $p(\la)q(\la)=\tp(\mu)\tq(\mu)$, from
which, after letting $\la,\mu\rightarrow\infty$ one obtains
\begin{equation}
 -2\mF_{t_1,\tilde{t}_1} = -2\mF_{\tilde{t}_1,t_1} \equiv u,
\label{Fu}
\end{equation}
where $u=u(t_1,t_2,\ldots;\tilde{t}_1,\tilde{t}_2,\ldots)$ is a
scalar function. Morever, for arbitrary $\la$, one has
\begin{equation}
p(\la)q(\la)=\tp(\la)\tq(\la)=u.  \label{dvn}
\end{equation}
Denoting $H_{2n+1}=2\pa_{t_{2n+1}}\pa_{t_1}\mF$,
$\hat{H}_{2n+1}=2\pa_{t_{2n+1}}\pa_{\tilde{t}_1}\mF$,
$\tilde{H}_{2n+1}=2\pa_{\tilde{t}_{2n+1}}\pa_{\tilde{t}_1}\mF$ and
$\tilde{\hat{H}}_{2n+1}=2\pa_{\tilde{t}_{2n+1}}\pa_{t_1}\mF$, the
Eq. (\ref{Fu}) can be treated as the evolution of $u$ with respect
to $t_{2n+1}$ and $\tilde{t}_{2n+1}$, respectively, namely
\begin{eqnarray}
 \frac{\pa u}{\pa t_{2n+1}} &=& -(H_{2n+1})_{\tilde{t}_1} = -(\hat{H}_{2n+1})_{t_1},
\label{u-t}\\
 \frac{\pa u}{\pa\tilde{t}_{2n+1}} &=& -(\tilde{H}_{2n+1})_{t_1}=-(\tilde{\hat{H}}_{2n+1})_{\tilde{t}_1}.
\label{u-tt}
\end{eqnarray}
Now we define the $\tau_{2n+1}$-flow by setting
$\pa_{\tau_{2n+1}}=\pa_{t_{2n+1}}+\pa_{\tilde{t}_{2n+1}}$
and identify $\tilde{t}_{2n+1}$ as the complex conjugate of $t_{2n+1}$,
in particular, $t_1:=z$ and $\tilde{t}_1:=\bz$, where $z=x+iy$.
From now on, the functions $\tilde{H}_{2n+1}$, $\tilde{\hat{H}}_{2n+1}$ can also be taken
as the complex conjugate of $H, \hat{H}$:
\[
 \tilde{H}_{2n+1}=\bar{H}_{2n+1}, \quad \tilde{\hat{H}}_{2n+1} = \bar{\hat{H}}_{2n+1}.
\]
Thus, incorporating (\ref{u-t}) and (\ref{u-tt}) for $n\geq 1$ together with
$n=0$ in (\ref{u-t}) (or (\ref{u-tt})) we obtain $\tau_{2n+1}$-flow of $u$
\begin{equation}
u_{\tau_{2n+1}} = -(H_{2n+1})_{\bz} -(\bar{H}_{2n+1})_{z}
 = -(\hat{H}_{2n+1})_{z} -(\bar{\hat{H}}_{2n+1})_{\bz}, \quad
u_z= -(H_1)_{\bz},
\label{u-tau}
\end{equation}
which is what we call the dVN hierarchy.
As in what follows, we shall show that for $n=1$,
the corresponding equation reduces to the dVN equation (\ref{dVNeq}).

%
\section{Faber polynomials of 2-dBKP}\label{Sec-Faber}
%
According to Takasaki's observations \cite{Tak06},
the left hand side of (\ref{EdBKP1}) (or (\ref{EdBKP3})) can be expanded as the form
\begin{equation}
 \log\frac{p(\la)-w}{p(\la)+w} = -\sum_{n=0}^{\infty}\frac{2\Phi_{2n+1}(w)}{2n+1}\la^{-2n-1},
\label{HF1}
\end{equation}
where $w=p(\mu)$ in (\ref{EdBKP1}) (or $w=\tilde{q}(\mu)$ in (\ref{EdBKP3}))
and $\Phi_n(w)$ is the $n$-th Faber polynomial of $p(\la)$ defined by
\begin{equation}
\log\frac{p(\la)-w}{\la} = -\sum_{n=1}^{\infty}\frac{\Phi_n(w)}{n}\la^{-n}.
\label{Faber}
\end{equation}
Eq. (\ref{Faber}) is analytic for large $\la$ for fixed $w\in\mathbb{C}$
and, (\ref{HF1}) is obtained from (\ref{Faber}) by taking into account the symmetry conditions
\[
 p(-\la) = -p(\la), \qquad \Phi_n(-w)=(-1)^n\Phi_n(w).
\]
If we replace $w$ in (\ref{HF1}) by $p(\mu)$, then Eq. (\ref{EdBKP1}) can be
reduced to the following system of Hamilton-Jacobi equation
\begin{equation}
 \frac{\pa S(\mu)}{\pa t_{2n+1}} = \Phi_{2n+1}(p(\mu)).
\label{HJ1}
\end{equation}
Likewise, replacing $w$ by $\tilde{q}(\mu)$, Eq. (\ref{EdBKP3}) reduces to
the following Hamilton-Jacobi equation
\begin{equation}
 \frac{\pa \tilde{S}(\mu)}{\pa t_{2n+1}} = \Phi_{2n+1}(\tilde{q}(\mu)).
\label{HJ2}
\end{equation}
After differentiating Eqs. (\ref{HJ1}), (\ref{HJ2}) with respect to $z, \bz$,
we have time evolutions of $p(\mu), q(\mu), \tilde{p}(\mu)$ and $\tilde{q}(\mu)$
in $t_{2n+1}$-flow in the following form
\begin{eqnarray}
&& \frac{\pa p(\mu)}{\pa t_{2n+1}} = \pa_{z}\Phi_{2n+1}(p(\mu)), \quad
   \frac{\pa q(\mu)}{\pa t_{2n+1}} = \pa_{\bz}\Phi_{2n+1}(p(\mu)),
\label{HJ-tpq1}\\
&& \frac{\pa \tilde{p}(\mu)}{\pa t_{2n+1}} = \pa_{\bz}\Phi_{2n+1}(\tilde{q}(\mu)), \quad
   \frac{\pa \tilde{q}(\mu)}{\pa t_{2n+1}} = \pa_{z}\Phi_{2n+1}(\tilde{q}(\mu)).
\label{HJ-tpq2}
\end{eqnarray}
In the same way, for the Hirota equations (\ref{EdBKP2}) and (\ref{EdBKP4}),
one can derive the corresponding Hamilton-Jacobi equations via the expression
of Faber polynomials as
\begin{equation}
 \log\frac{\tilde{p}(\la)-w}{\tilde{p}(\la)+w}
= -\sum_{n=0}^{\infty}\frac{2\tilde{\Phi}_{2n+1}(w)}{2n+1}\la^{-2n-1}.
\label{HF2}
\end{equation}
From which, substitutions of $w=\tilde{p}(\mu)$ and $w=q(\mu)$ yielding
the following systems of Hamilton-Jacobi equations
\[
\frac{\pa \tilde{S}(\mu)}{\pa \tilde{t}_{2n+1}} = \tilde{\Phi}_{2n+1}(\tilde{p}(\mu)), \quad
\frac{\pa S(\mu)}{\pa \tilde{t}_{2n+1}} = \tilde{\Phi}_{2n+1}(q(\mu)).
\]
Therefore, we have the following time evolutions of $p(\mu), q(\mu), \tilde{p}(\mu)$ and $\tilde{q}(\mu)$
with respect to $\tilde{t}_{2n+1}$-flow
\begin{eqnarray}
&& \frac{\pa \tilde{p}(\mu)}{\pa \tilde{t}_{2n+1}} = \pa_{\bz}\tilde{\Phi}_{2n+1}(\tilde{p}(\mu)), \quad
   \frac{\pa \tilde{q}(\mu)}{\pa \tilde{t}_{2n+1}} = \pa_{z}\tilde{\Phi}_{2n+1}(\tilde{p}(\mu)),
\label{HJ-ttpq1}\\
&& \frac{\pa p(\mu)}{\pa \tilde{t}_{2n+1}} = \pa_{z}\tilde{\Phi}_{2n+1}(q(\mu)), \quad
   \frac{\pa q(\mu)}{\pa \tilde{t}_{2n+1}} = \pa_{\bz}\tilde{\Phi}_{2n+1}(q(\mu)).
\label{HJ-ttpq2}
\end{eqnarray}

To see how the Faber polynomials will generate functions
$H_{2n+1},\hat{H}_{2n+1}, \tilde{H}_{2n+1}$ and $\tilde{\hat{H}}_{2n+1}$
shown in Eqs. (\ref{u-t}), (\ref{u-tt}),
similar derivations in Teo's paper \cite{Teo03} (see also \cite{Pom75}),
we differentiate (\ref{HF1}) to the both sides with respect to $\la$ and obtain
\[
 \frac{w p'(\la)}{p^2(\la)-w^2}=\sum_{n=0}^{\infty}\Phi_{2n+1}(w)\la^{-2n-2}.
\]
Putting
$p(\la)=\la-\sum_{n=0}^{\infty}\frac{H_{2n+1}}{2n+1}\la^{-2n-1}$
into this expression, then we have
\[
 w+w\sum_{n=0}^{\infty}H_{2n+1}\la^{-2n-2}
=\left[\left(\la-\sum_{n=0}^{\infty}\frac{H_{2n+1}}{2n+1}\la^{-2n-1}\right)^2 - w^2\right]
 \left(\sum_{n=0}^{\infty}\Phi_{2n+1}(w)\la^{-2n-2}\right).
\]
Comparing coefficients of all powers of $\la$ on both sides, we have
\begin{eqnarray}
 \Phi_1(w) &=& w, \no\\
 \Phi_3(w) &=& w^3 + 3H_1w
\label{P1-P3}
\end{eqnarray}
and the recursion formula
\begin{eqnarray*}
 \Phi_{2n+5}(w)
&=& w^2\Phi_{2n+3}(w)
  - \sum_{m=0}^n\sum_{k=0}^{n-m}\frac{H_{2n-2m-2k+1}H_{2k+1}}{(2n-2m-2k+1)(2k+1)}\Phi_{2m+1}(w) \\
  && + 2\sum_{m=0}^{n+1}\frac{H_{2n-2m+3}}{2n-2m+3}\Phi_{2m+1}
  + wH_{2n+3}, \qquad n=0,1,2,\ldots,
\end{eqnarray*}
which can be used to solve for $\Phi_n$.
The first few of $\Phi_n(w)$ are given by
\begin{eqnarray}
\Phi_5(w) &=& w^5+5H_1w^3+5(H_1^2+H_3/3)w, \no\\
\Phi_7(w) &=& w^7+7H_1w^5+7(2H_1^2+H_3/3)w^3+7(H_1^3+(2/3)H_1H_3+H_5/5)w, \no\\
\Phi_9(w) &=& w^9+9H_1w^7+9(3H_1^2+H_3/3)w^5+3(10H_1^3+4H_1H_3+(3/5)H_5)w^3 \no\\
          &&  +9(H_1^4+H_1^2H_3+H_3^2/9+(2/5)H_1H_5+H_7/7)w.
\label{P5-P9}
\end{eqnarray}
Similarly, differentiating (\ref{HF2}) with respect to $\la$, we have
\[
 \frac{w \tilde{p}'(\la)}{\tilde{p}^2(\la)-w^2}=\sum_{n=0}^{\infty}\tilde{\Phi}_{2n+1}(w)\la^{-2n-2}.
\]
Now putting $\tilde{p}(\la)=\la-\sum_{n=0}^{\infty}\frac{\bar{H}_{2n+1}}{2n+1}\la^{-2n-1}$
into this expression and comparing coefficients of powers of $\la$,
we derive the first few expressions of Faber polynomials
\begin{eqnarray}
\tilde{\Phi}_1(w) &=& w, \no\\
\tilde{\Phi}_3(w) &=& w^3 + 3\bar{H}_1w, \no\\
\tilde{\Phi}_5(w) &=& w^5+5\bar{H}_1w^3+5(\bar{H}_1^2+\bar{H}_3/3)w, \no\\
\tilde{\Phi}_7(w) &=& w^7+7\bar{H}_1w^5+7(2\bar{H}_1^2+\bar{H}_3/3)w^3
                      +7(\bar{H}_1^3+(2/3)\bar{H}_1\bar{H}_3+\bar{H}_5/5)w, \no\\
\tilde{\Phi}_9(w) &=& w^9+9\bar{H}_1w^7+9(3\bar{H}_1^2+\bar{H}_3/3)w^5
                      +3(10\bar{H}_1^3+4\bar{H}_1\bar{H}_3+(3/5)\bar{H}_5)w^3 \no\\
&&          +9(\bar{H}_1^4+\bar{H}_1^2\bar{H}_3+\bar{H}_3^2/9
               +(2/5)\bar{H}_1\bar{H}_5+\bar{H}_7/7)w, \no\\
&\vdots &
\label{tP1-tP9}
\end{eqnarray}
in which $\tilde{\Phi}_{2n+1}$ obey the recurrence relations
\begin{eqnarray*}
 \tilde{\Phi}_{2n+5}(w)
&=& w^2\tilde{\Phi}_{2n+3}(w)
  - \sum_{m=0}^n\sum_{k=0}^{n-m}\frac{\bar{H}_{2n-2m-2k+1}\bar{H}_{2k+1}}
                                     {(2n-2m-2k+1)(2k+1)}\tilde{\Phi}_{2m+1}(w) \\
  && + 2\sum_{m=0}^{n+1}\frac{\bar{H}_{2n-2m+3}}{2n-2m+3}\tilde{\Phi}_{2m+1}
  + w\bar{H}_{2n+3}, \qquad n=0,1,2,\ldots.
\end{eqnarray*}
%
\section{Symmetry constraint of dVN hierarchy and conserved densities}\label{Sec-sym}
%
One way of determine the conserved densities of the dVN hierarchy
is to impose the desired symmetry constraints \cite{BKM06}, so
that the explicit formulas of these densities can be connected to
the corresponding Faber polynomials and be solved recursively. The
main symmetry constraint we considere is of the form
\begin{equation}
 u_x=(S^i)_{z\bar{z}},
\label{sym-con-1}
\end{equation}
where $S^i=S(\mu_i)$ is evaluated at some point $\mu_i$ and we
assume $S^i$ is real number. In this section, we would like to
show that all of the conserved densities can be derived by means
of the associated Faber polynomials under this symmetry reduction.
We discuss these relations along the following two ways.
\\
\textbf{(I)}
Let us take the derivatives of $S(\la)$ with respect to $z, \bz,
x$ and, noticing that $-2\mF_{z\bz}=u$, we have
\begin{equation}
 \frac{\pa^3S(\la)}{\pa z\pa\bar{z}\pa x}
= -D(\la)\mF_{z\bar{z}x}= \frac{1}{2}D(\la)u_x=\frac{1}{2}D(\la)S^i_{z\bar{z}},
\label{D3_S}
\end{equation}
in which $\pa S^i/\pa z=p(\mu_i)=p^i$,  $\pa S^i/\pa
\bar{z}=q(\mu_i)=q^i= \bar (\pa S^i/\pa z)=\bar p^i$ obey the
algebraic relation $u=p^iq^i=p^i \bar p^i$ and then $u$ is
positive  real number. We remark here that in the context of
nonlinear geometry optics, the quantity $\sqrt{u}$ is proportion
to the refractive index and $u=S^i_zS^i_{\bz}$ is nothing but the
standard Eikonal equation arises from the high-frequency limit of
Maxwell equations \cite{BKM06,KM04a,KM05,KM04b}. \\
\indent Integrating (\ref{D3_S}) with respect to $z,\bar{z}$
respectively and considering (\ref{EdBKP1}), it follows that
\begin{eqnarray}
(p(\la))_x &=& \left(\frac{1}{2}D(\la)S^i\right)_z
            = -\frac{1}{2}\pa_z\left(\log\frac{p(\la)-p^i}{p(\la)+p^i}\right),
\label{p_x}\\
(q(\la))_x &=& \left(\frac{1}{2}D(\la)S^i\right)_{\bar{z}}
            = -\frac{1}{2}\pa_{\bar{z}}\left(\log\frac{p(\la)-p^i}{p(\la)+p^i}\right).
\label{bp_x}
\end{eqnarray}
Using (\ref{HF1}) with $w$ replaced by $p^i$ and the expansions of $p(\la)$ and $q(\la)$,
Eqs. (\ref{p_x}) and (\ref{bp_x}) can be rewritten respectively by
\begin{eqnarray}
 \pa_xH_{2n+1} &=& -\pa_z\Phi_{2n+1}(p^i),
\label{Hamil-Faber} \\
 \pa_x\hat{H}_{2n+1} &=& -\pa_{\bar{z}}\Phi_{2n+1}(p^i),
\label{hHamil-Faber}
\end{eqnarray}
where $H_{2n+1}\equiv 2\pa_{z}\pa_{t_{2n+1}}\mF$ and
$\hat{H}_{2n+1}\equiv 2\pa_{\bar{z}}\pa_{t_{2n+1}}\mF$. Hence,
Eqs. (\ref{Hamil-Faber}), (\ref{hHamil-Faber}) provide the
Hamilton-Jacobi equations (\ref{HJ-tpq1}), which can now be read
as
\begin{equation}
\frac{\pa p^i}{\pa t_{2n+1}} = -\frac{\pa H_{2n+1}}{\pa x}, \quad
\frac{\pa \bp^i}{\pa t_{2n+1}} = -\frac{\pa\hat{H}_{2n+1}}{\pa x}.
\label{t-2n+1}
\end{equation}
As the result, the functions $H_{2n+1}$ and $\hat{H}_{2n+1}$ appear to be the
\emph{conserved densities} that characterized by the associated Hamilton-Jabobi equations.
Furthermore, from (\ref{Hamil-Faber}), (\ref{hHamil-Faber})
we see that $H_{2n+1}$ and $\hat{H}_{2n+1}$ are related by the compatibility relations
\[
 \pa_{\bar{z}}H_{2n+1} = \pa_z\hat{H}_{2n+1},
\]
and can be obtained by solving Eqs. (\ref{Hamil-Faber}) and (\ref{hHamil-Faber}).
\\
\textbf{(II)}
In the similar way, the differentiation of $\tilde{S}(\la)$ with respect to $z, \bz, x$ shows that
\[
 \frac{\pa^3\tilde{S}(\la)}{\pa z\pa\bar{z}\pa x}
= -\tD(\la)\mF_{z\bar{z}x}= \frac{1}{2}\tD(\la)u_x=\frac{1}{2}\tD(\la)S^i_{z\bar{z}},
\]
Then we get
\begin{eqnarray}
(\tp(\la))_x &=& \left(\frac{1}{2}\tD(\la)S^i\right)_{\bz}
            = -\frac{1}{2}\pa_{\bz}\left(\log\frac{\tp(\la)-\bp^i}{\tp(\la)+\bp^i}\right),
\label{tp_x}\\
(\tq(\la))_x &=& \left(\frac{1}{2}\tD(\la)S^i\right)_{z}
            = -\frac{1}{2}\pa_{z}\left(\log\frac{\tp(\la)-\bp^i}{\tp(\la)+\bp^i}\right).
\label{tbp_x}
\end{eqnarray}
Using (\ref{HF2}) with $w$ replaced by $\bp^i$ and the expansion of $\tp(\la)$,
we rewrite (\ref{tp_x}) and (\ref{tbp_x}) as
\begin{eqnarray}
 \pa_x\bar{H}_{2n+1} &=& -\pa_{\bz}\tilde{\Phi}_{2n+1}(\bp^i),
\label{tHamil-Faber} \\
 \pa_x\bar{\hat{H}}_{2n+1} &=& -\pa_{z}\tilde{\Phi}_{2n+1}(\bp^i).
\label{htHamil-Faber}
\end{eqnarray}
where $\bar{H}_{2n+1}\equiv 2\pa_{\bz}\pa_{\bt_{2n+1}}\mF$ and
$\bar{\hat{H}}_{2n+1}\equiv 2\pa_{z}\pa_{\bt_{2n+1}}\mF$.
Therefore, the Hamilton-Jacobi equations (\ref{HJ-ttpq2}) can now be read by
the conservation laws:
\begin{equation}
\frac{\pa \bp^i}{\pa \bt_{2n+1}} = -\frac{\pa\bar{H}_{2n+1}}{\pa x}, \quad
\frac{\pa p^i}{\pa \bt_{2n+1}} = -\frac{\pa\bar{\hat{H}}_{2n+1}}{\pa x}.
\label{tt-2n+1}
\end{equation}
Also, the conserved densities $\bar{H}_{2n+1}$ and
$\bar{\hat{H}}_{2n+1}$ in (\ref{tHamil-Faber}) and
(\ref{htHamil-Faber}) satisfy the compatibilities
\[
 \pa_{z}\bar{H}_{2n+1} = \pa_{\bz}\bar{\hat{H}}_{2n+1}
\]
and can be solved according to (\ref{tHamil-Faber}), (\ref{htHamil-Faber}).
Notice that the Faber polynomials $\tilde{\Phi}_{2n+1}(\bp^i)$
have become the complex conjugate of $\Phi_{2n+1}(p^i)$
i.e., $\tilde{\Phi}_{2n+1}(\bp^i)=\overline{\Phi_{2n+1}(p^i)}$.
We remark here that since $u=p^i\bp^i$ in those $H_n$'s,
$\Phi_{2n+1}$ is understood as functions of $p^i,\bp^i$.

In the following, under the symmetry constraint (\ref{sym-con-1})
we shall give some examples to demonstrate how to solve
conserved densities $H_{2n+1}$, $\hat{H}_{2n+1}$.
Then, $\bar{H}_{2n+1}$ and $\bar{\hat{H}}_{2n+1}$ are given automatically
by taking the complex conjugate of $H_{2n+1}$ and $\hat{H}_{2n+1}$, respectively.
For simplifying calculations, we shall use Faber polynomials
(\ref{P1-P3}), (\ref{P5-P9}) and the useful identities
\begin{eqnarray}
 p^i_z &=& p^i_x-u_x, \label{id1}\\
u_zp^i &=& up^i_x-uu_x+u_x(p^i)^2.
\label{id2}
\end{eqnarray}
We determine the relationship of $\hat{H}_{2n+1}$ and $H_{2n+1}$
from the  equation \eqref{dvn}. Noting that  $p(\la)$ and $q(\la)$
are defined by
\[
p(\la) = \la - \sum_{n=0}^{\infty}\frac{H_{2n+1}}{2n+1}\la^{-2n-1}, \quad
q(\la) = -\sum_{n=0}^{\infty}\frac{\hat{H}_{2n+1}}{2n+1}\la^{-2n-1},
\]
and putting $p(\la)$ and $q(\la)$ into \eqref{dvn}, we have the
expression:
\begin{equation}
 u = -\sum_{n=0}^{\infty}\frac{\hat{H}_{2n+1}}{2n+1}\la^{-2n}
     +\sum_{n,m=0}^{\infty}\frac{H_{2n+1}\hat{H}_{2m+1}}{(2n+1)(2m+1)}\la^{-2n-2m-2}.
\end{equation}
Identifying the coefficients of all powers of $\la$ at the both
sides, we obtain
\[
 \hat{H}_1 = -u
\]
and the recursion relation of $\hat{H}_{2n+1}$ and $H_{2n+1}$ by
\begin{equation}
 \hat{H}_{2n+3} = (2n+3)\sum_{k=0}^n\frac{H_{2n-2k+1}\hat{H}_{2k+1}}{(2n-2k+1)(2k+1)},
 \qquad n=0,1,2,\ldots.
\label{recur2}
\end{equation}
Some of them are given by
\begin{eqnarray*}
 \hat{H}_3 &=& -3uH_1, \\
 \hat{H}_5 &=& -\frac{5}{3}u(3H_1^2+H_3), \\
 \hat{H}_7 &=& -\frac{7}{3}u(3H_1^3+2H_1H_3+\frac{3}{5}H_5).
\end{eqnarray*}
\\[0.3cm]
\textbf{Examples of constructing conserved densities}
\\[0.3cm]
%
\textit{Example 1}.
By (\ref{Hamil-Faber}), for $n=0$
\[
 H_{1x}=-\Phi_{1z}=-p^i_z=(u-p^i)_x,
\]
where we have used (\ref{id1}). Integrating both sides with
respect to $x$ yields
\begin{equation}
 H_1=u-p^i.
\label{H1}
\end{equation}
\textit{Example 2}.
 By (\ref{Hamil-Faber}), for $n=1$,
\begin{eqnarray*}
 H_{3x} &=& -\Phi_{3z}
         =  -\left((p^i)^3+3(u-p^i)p^i\right)_z, \\
        &=& \left(3(u-p^i)^2-(p^i)^3\right)_x
         =  \left(3H_1^2-(p^i)^3\right)_x,
\end{eqnarray*}
where we have used Eqs. (\ref{P1-P3}), (\ref{H1}) in the first
line and (\ref{id1}), (\ref{id2}) to obtain the second line. After
integrating both sides with respect to $x$, we get
\begin{equation}
H_3=3H_1^2-(p^i)^3.
\label{H3}
\end{equation}
\textit{Example 3}.
For $n=2$ in (\ref{Hamil-Faber}), using (\ref{P5-P9}), (\ref{H1}), (\ref{H3})
and the identities (\ref{id1}), (\ref{id2}) we have
\begin{eqnarray*}
 H_{5x} &=& -\Phi_{5z}
         =  -\left((p^i)^5+5(u-p^i)(p^i)^3+5\left(2(u-p^i)^2-\frac{1}{3}(p^i)^3\right)p^i\right)_z, \\
        &=& \left(10(u-p^i)^3-\frac{20}{3}(u-p^i)(p^i)^3-(p^i)^5\right)_x
         =  \left(10H_1^3-\frac{20}{3}H_1(p^i)^3-(p^i)^5\right)_x, \\
        &=& \left(-10H_1^3+\frac{20}{3}H_1H_3-(p^i)^5\right)_x,
\end{eqnarray*}
where we have used the substitution for $(p^i)^3$ by (\ref{H3}) to
obtain the last equality. Integrating both sides with respect to
$x$, we have
\begin{equation}
H_5=-10H_1^3+\frac{20}{3}H_1H_3-(p^i)^5.
\label{H5}
\end{equation}
\textit{Example 4}.
For $n=3$, similarly, we have
\begin{eqnarray*}
 H_{7x}
&=& -\Phi_{7z}, \\
&=& -\left\{(p^i)^7+7(u-p^i)(p^i)^5+7\left(3(u-p^i)^2-\frac{1}{3}(p^i)^3\right)(p^i)^3 \right. \\
&&   +7\left[(u-p^i)^3+2(u-p^i)\left((u-p^i)^2-\frac{1}{3}(p^i)^3\right) \right. \\
&& \left.\left.-2(u-p^i)^3+4(u-p^i)\left((u-p^i)^2-\frac{1}{3}(p^i)^3\right)
               -\frac{1}{5}(p^i)^5 \right]p^i\right\}_z,\\
&=& 7\left(5(u-p^i)^4-5(u-p^i)^2(p^i)^3-\frac{6}{5}(u-p^i)(p^i)^5
           +\frac{1}{3}(p^i)^6-\frac{1}{7}(p^i)^7\right)_x.
\end{eqnarray*}
Again, with substitutions for $(p^i)^3$ and $(p^i)^5$ obtained by
(\ref{H3}) and (\ref{H5}) respectively and integrating over $x$,
we solve
\begin{equation}
H_7=7\left(5H_1^4-5H_1^2H_3+\frac{6}{5}H_1H_5+\frac{1}{3}H_3^2-(p^i)^7/7\right).
\label{H7}
\end{equation}

%
\section{Hodograph solutions of dVN hierarchy}\label{Sec-hodo}
%
Having set up the Faber polynomials in terms of $p^i,\bp^i$ for
the dVN hierarchy that underline the imposed symmetry constraint
\eqref{sym-con-1}, now we would like to use the hodograph method
to find  the hodograph solutions of $p^i(z,\bz,\tau_{2n+1})$ and
$\bp^i(z,\bz,\tau_{2n+1})$. Hence we shall obtain solutions of the
dVN equation.

From (\ref{Hamil-Faber})-(\ref{t-2n+1}) and (\ref{tHamil-Faber})-(\ref{tt-2n+1}),
the $t_{2n+1}$- and $\bt_{2n+1}$-flows of $p^i, \bp^i$
can be written respectively by
\begin{eqnarray*}
\begin{pmatrix} p^i \\ \bp^i \end{pmatrix}_{t_{2n+1}}
&=&
\pa_{p^i}\Phi_{2n+1}\begin{pmatrix}p^i \\ \bp^i\end{pmatrix}_z
+\pa_{\bp^i}\Phi_{2n+1}\begin{pmatrix}p^i \\ \bp^i\end{pmatrix}_{\bz}, \\
\begin{pmatrix} p^i \\ \bp^i \end{pmatrix}_{\bt_{2n+1}}
&=&
\pa_{p^i}\overline{\Phi_{2n+1}}\begin{pmatrix}p^i \\ \bp^i\end{pmatrix}_z
+\pa_{\bp^i}\overline{\Phi_{2n+1}}\begin{pmatrix}p^i \\ \bp^i\end{pmatrix}_{\bz},
\end{eqnarray*}
where we have used the fact that $p^i_{\bz}=\bp^i_z$.
Therefore, the $\tau_{2n+1}$-flow of the dVN hierarchy is governed by
\begin{equation}
\begin{pmatrix} p^i \\ \bp^i \end{pmatrix}_{\tau_{2n+1}}=
\pa_{p^i}M_{2n+1}\begin{pmatrix}p^i \\ \bp^i\end{pmatrix}_z
+\pa_{\bp^i}M_{2n+1}\begin{pmatrix}p^i \\ \bp^i\end{pmatrix}_{\bz}, \quad n\geq 1
\label{tau-flow}
\end{equation}
where $M_{2n+1}\equiv\Phi_{2n+1}+\overline{\Phi_{2n+1}}$.
Note that the first equation $n=0$ of the hierarchy says that $p^i$ and $\bp^i$ depend on $\tau_1$ and $x$
only through the linear combination $\tau_1+x$.
It is easy to see that the above equation has the following implicit form of hodograph equations
\begin{align}
\begin{split}
 & z+\sum_{n=1}^{\infty}f_{2n+1}(p^i,\bp^i)\tau_{2n+1} = F(p^i,\bp^i), \\
 & \bz+\sum_{n=1}^{\infty}g_{2n+1}(p^i,\bp^i)\tau_{2n+1} = G(p^i,\bp^i),
\end{split}
\label{hodo-tau}
\end{align}
where $F$ and $G$ are the initial data at $\tau_{2n+1}=0$, and
 $f_{2n+1}=\pa_{p^i}M_{2n+1},\; g_{2n+1}=\pa_{\bp^i}M_{2n+1}$.
One can show that, because of commutativity of the $\tau_{2n+1}$-flows of $p^i, \bp^i$,
$G$ and $F$ obey the following constraints
\begin{eqnarray}
F_{\bp^i} &=& G_{p^i}, \label{FG-1}\\
p^iF_{p^i} &=& -\bp^i G_{\bp^i} - (1-p^i-\bp^i)G_{p^i}. \label{FG-2}
\end{eqnarray}
It turns out that in (\ref{FG-1}) there exists a function
$\vphi(p^i,\bp^i)$ such that $F=\pa_{p^i}\vphi$, $G=\pa_{\bp^i}\vphi$.
Substituting into (\ref{FG-2}) we have the defining equation for $\vphi$
\begin{equation}
p^i\vphi_{p^ip^i}+\bp^i\vphi_{\bp^i\bp^i}+(1-p^i-\bp^i)\vphi_{p^i\bp^i} =0.
\label{phi-eq}
\end{equation}
Let $p^i=(\rho_1-i\rho_2)/2$, $\bp^i=(\rho_1+i\rho_2)/2$, we have
$\pa/\pa p^i=\pa/\pa\rho_1-i\pa/\pa\rho_2$ and $\pa/\pa\bp^i=\pa/\pa\rho_1+i\pa/\pa\rho_2$.
Then the defining equation (\ref{phi-eq}) becomes
\begin{equation}
 \vphi_{\rho_1\rho_1} + 2\rho_2\vphi_{\rho_1\rho_2} + (1-2\rho_1)\vphi_{\rho_2\rho_2} = 0.
\label{phi-eq2}
\end{equation}
In fact, due to the existence of $\vphi$, the functions $F, G$ can
be chosen as a natural setting in the linear combination of $f$
and $g$ defined by (\ref{hodo-tau}). Namely, $F=\sum_{n\geq
0}\mu_nf_{2n+1}$ and $G=\sum_{n\geq 0}\xi_ng_{2n+1}$ with
constraint $\mu_n=\xi_n$. We deduce that $\vphi$ has the
polynomial type expansion in $\rho_1, \rho_2$:
\begin{equation}
\vphi=\sum_{n=0}^{\infty}\mu_nM_{2n+1}(\rho_1,\rho_2)
\label{phi-poly}
\end{equation}
satisfies (\ref{phi-eq2}).
For instance, some cases are established as follows.
\\
\textbf{(i)}
$\vphi=M_1=\Phi_1+\overline{\Phi_1}=\rho_1$. It is obvious. \\
\textbf{(ii)}
$\vphi=M_3=\Phi_3+\overline{\Phi_3}=\rho_1^3-3\rho_1^2+\frac{3}{2}(\rho_1^2+\rho_2^2)$.
\\
 \textbf{(iii)} $\vphi=M_5=\Phi_5+\overline{\Phi_5}
=\rho_1^5-\frac{20}{3}\rho_1^4+10\rho_1^3+\frac{5}{3}\rho_1^2(\rho_1^2+\rho_2^2)
 -\frac{15}{2}\rho_1(\rho_1^2+\rho_2^2)
 +\frac{5}{3}(\rho_1^2+\rho_2^2)^2$. \\
\textbf{(iv)} $\vphi=M_7=\Phi_7+\overline{\Phi_7}$ has expression
as
\begin{eqnarray*}
\vphi
&=& \rho_1^{7}-{\frac{259}{30}}\,\rho_1^{6}+{\frac {35}{2}}\,\rho_1^{5}
   +{\frac {21}{10}}\,\rho_1^{4}\rho_2^{2}+\frac{7}{5}\,(\rho_1^2+\rho_2^2)^{2}\rho_1^{2}
   -{\frac{35}{2}}\,\rho_1^{3}\rho_2 ^{2} \\
&& +35\,\rho_1^{2}\rho_2^{2}-{\frac{35}{4}}\,(\rho_1^2+\rho_2^2)^{2}\rho_1
  -{\frac {35}{8}}\,(\rho_1^2+\rho_2^2) ^{2} +{\frac{28}{15}}\,(\rho_1^2+\rho_2^2)^{3}.
\end{eqnarray*}
\textbf{Remark.} One can also find several simple solutions of the
certain PDEs in Eq. (\ref{phi-eq2}) in the following ways:
(a) $\vphi_{\rho_1\rho_1}=0$, $2\rho_2\vphi_{\rho_1\rho_2}+(1-2\rho_1)\vphi_{\rho_2\rho_2}=0$,
(b) $\vphi_{\rho_1\rho_2}=0$, $\vphi_{\rho_1\rho_1}+(1-2\rho_1)\vphi_{\rho_2\rho_2} =0$,
and (c) $\vphi_{\rho_2\rho_2}=0$, $\vphi_{\rho_1\rho_1}+2\rho_2\vphi_{\rho_1\rho_2}=0$.
It can be shown that cases (a) and (b) have solutions of polynomial type involved in (\ref{phi-poly}),
while (c) is not the case. For example, case (c) has solutions of the form:
$\vphi=c_0+c_1\rho_1+c_2\rho_2+c_3\rho_2\exp(-2\rho_1)$.
%
%
%
\subsection*{Example}
To find the (2+1)-dimensional solutions involving $(z,\bz,\tau)$ that satisfy (\ref{tau-flow}),
using $\Phi_3=(p^i)^3+3H_1p^i$ where $H_1=p^i\bp^i-p^i$, we expand the hodograph equation
(\ref{hodo-tau}) up to $\tau_3=\tau$:
\begin{align*}
\begin{split}
 F(p^i,\bp^i)&= z+f_3\,\tau = z+ \bigg(3(p^i+\bp^i)^2-6p^i\bigg)\tau , \\
 G(p^i,\bp^i)&= \bz+g_3\,\tau =\bz+ \bigg(3(p^i+\bp^i)^2-6\bp^i\bigg)\tau.
\end{split}
\end{align*}
Choosing $F=1, G=1$, the above equations can be easily solved by
\begin{eqnarray*}
p^i &=& \frac{1}{12\tau}\left(3\tau+(z-\bz)\pm\sqrt{9\tau^2-6\tau(z+\bz-2)}\right), \\
\bp^i &=& \frac{1}{12\tau}\left(3\tau-(z-\bz)\pm\sqrt{9\tau^2-6\tau(z+\bz-2)}\right).
\end{eqnarray*}
Then $u$ is read as
\begin{equation}
u = p^i\bp^i
  = \frac{1}{144\tau^2}\bigg(18\tau^2-6\tau(z+\bz-2)-(z-\bz)^2
    \pm 6\tau\sqrt{9\tau^2-6\tau(z+\bz-2)}\bigg).
\label{hodo-sol-1}
\end{equation}
One can verify that (\ref{hodo-sol-1}) satisfies the
 dVN equation (\ref{dVNeq}) with $\tau=\tau_3, V=3 H_1$.
Furthermore, if we choose $F=f_3, G=g_3$, we get
\begin{eqnarray*}
p^i &=& \frac{1}{12(\tau-1)}\left(3(\tau-1)+(z-\bz)\pm\sqrt{9(\tau-1)^2-6(\tau-1)(z+\bz)}\right), \\
\bp^i &=& \frac{1}{12(\tau-1)}\left(3(\tau-1)-(z-\bz)\pm\sqrt{9(\tau-1)^2-6(\tau-1)(z+\bz)}\right).
\end{eqnarray*}
Therefore,
\[
u = \frac{18(\tau-1)^2-6(\tau-1)(z+\bz)-(z-\bz)^2
    \pm 6(\tau-1)\sqrt{9(\tau-1)^2-6(\tau-1)(z+\bz)}}{144(\tau-1)^2}.
\]
More new solutions can be given in this manner, but the main
difficulty we have to confront with is to solve higher order
algebraic equations. Finally, we want to solve $S^i$ function of
the above example via the partial differentiations $\pa S^i/\pa
z=p^i$ and $\pa S^i/\pa\bz=\bp^i$. It is easy to obtain that the
expression of $S^i$ is given by
\[
 S^i(z,\bz,\tau)
= \frac{3(\tau-1)(z-\bz)^2+18(\tau-1)^2(z+\bz+4C)
  -2\sqrt{3}\left(3(\tau-1)^2-2(\tau-1)(z+\bz)\right)^{3/2}}{72(\tau-1)^2},
\]
where $C$ is an arbitrary constant.

%
\section{$2N$-component case}\label{Sec-2N}
%
In this section, we  give an $2N$-component reduction of the dVN
hierarchy under a more general symmetry constraint, and construct
the corresponding hodograph equation. Let us consider the symmetry
constraint of the form \cite{BKM06}
\begin{equation}
 u_x = \sum_{i=1}^{N}\ep_i S^i_{z\bz}.
\label{sym-con-2}
\end{equation}
Particularly, we impose two assumptions: $u=p^i\bp^i$, $\forall\,i=1,\ldots,N$ and $\sum_{i=1}^N\ep_i=1$.
Similar calculations in Sec.\ref{Sec-sym}, we have the following relations between
conserved densities and the associated Faber polynomials:
\begin{eqnarray*}
  (H_{2n+1})_x &=& -\sum_{i=1}^N\ep_i\pa_z\Phi_{2n+1}(p^i), \\
  (\hat{H}_{2n+1})_x &=& -\sum_{i=1}^N\ep_i\pa_{\bz}\Phi_{2n+1}(p^i),
\end{eqnarray*}
where the Faber polynomials $\Phi_{2n+1}(p^i)$ are defined as
before, in which the  conserved densities have different forms and
can also be determined recursively. Some of $H_{2n+1}$ for the
$2N$-reduction system are given by
\begin{eqnarray*}
&& H_1=u-\sum_{i=1}^N\ep_ip^i, \quad
H_3=3H_1^2-\sum_{i=1}^N\ep_i(p^i)^3,\quad
H_5=-10H_1^3+\frac{20}{3}H_1H_3-\sum_{i=1}^N\ep_i(p^i)^5, \\
&&H_7=35H_1^4-35H_1^2H_3+\frac{42}{5}H_1H_5+\frac{7}{3}H_3^2-\sum_{i=1}^N\ep_i(p^i)^7.
\end{eqnarray*}
In terms of these $H_n$'s, the expressions of $\hat{H}_{2n+1}$ follow
the same as the presented form in (\ref{recur2}).
Under the symmetry constraint, the Hamilton-Jacobi equations can now be written in the
following way:
\begin{eqnarray*}
&&\frac{\pa p^k}{\pa t_{2n+1}} = \pa_z \Phi_{2n+1}(p^k;p^1,\ldots,p^N,\bp^1,\ldots,\bp^N), \quad
  \frac{\pa \bp^k}{\pa t_{2n+1}} = \pa_{\bz} \Phi_{2n+1}(p^k;p^1,\ldots,p^N,\bp^1,\ldots,\bp^N), \\
&&\frac{\pa p^k}{\pa \bt_{2n+1}} = \pa_z \bar{\Phi}_{2n+1}(\bp^k;p^1,\ldots,p^N,\bp^1,\ldots,\bp^N), \quad
  \frac{\pa \bp^k}{\pa \bt_{2n+1}} = \pa_{\bz} \bar{\Phi}_{2n+1}(\bp^k;p^1,\ldots,p^N,\bp^1,\ldots,\bp^N),
\end{eqnarray*}
where $k=1,\ldots,N$.
After incorporating the above evolution equations to the $\tau_{2n+1}$-flow of dVN hierarchy
and noting that $p^i_{\bz}=\bp^i_z$ for $i=1,\ldots,N$, we arrive the hodograph equation
of $2N$-component system
\begin{equation}
\begin{pmatrix}p^k \\ \bp^k\end{pmatrix}_{\tau_{2n+1}}
= \sum_{i=1}^Nf^i_{2n+1}
    \begin{pmatrix}p^i\\ \bp^i\end{pmatrix}_z
  +\sum_{i=1}^Ng^i_{2n+1}
    \begin{pmatrix}p^i\\ \bp^i\end{pmatrix}_{\bz},
\qquad k=1,\ldots,N, \quad n\geq 1,
\label{2N-hodo}
\end{equation}
where $f^i_{2n+1}(p^k,\bp^k)=\pa_{p^i}(\Phi_{2n+1}(p^k)+\bar{\Phi}_{2n+1}(\bp^k))$ and
$g^i_{2n+1}(p^k,\bp^k)=\pa_{\bp^i}(\Phi_{2n+1}(p^k)+\bar{\Phi}_{2n+1}(\bp^k))$.
For example, in the case of $N=2$ we have
\[
\begin{pmatrix} p^1 \\ \bp^1 \\ p^2 \\ \bp^2 \end{pmatrix}_{\tau_{2n+1}}
=\begin{pmatrix} \mbox{~} & \mbox{~} \\[-0.1cm]
                 f^1(p^1,\bp^1) I_2 + g^1(p^1,\bp^1)\mathbb{A}(p^1) &
                 f^2(p^1,\bp^1) I_2 + g^2(p^1,\bp^1)\mathbb{A}(p^2) \\[0.4cm]
                 f^1(p^2,\bp^2) I_2 + g^1(p^2,\bp^2)\mathbb{A}(p^1) &
                 f^2(p^2,\bp^2) I_2 + g^2(p^2,\bp^2)\mathbb{A}(p^2) \\[-0.2cm]
                 \mbox{~} & \mbox{~}
 \end{pmatrix}
\begin{pmatrix} p^1 \\ \bp^1 \\ p^2 \\ \bp^2 \end{pmatrix}_z,
\]
where $I_2$ is the $2\times 2$ identity matrix and
\[
 \mathbb{A}(p^i) =\begin{pmatrix} 0 & 1 \\ -\frac{\bp^i}{p^i} & \frac{1-p^i-\bp^i}{p^i}\end{pmatrix},
 \qquad i=1,2.
\]

%
\section{Concluding remarks}\label{Sec-Con}
%
In this paper we have studied dVN hierarchy from the framework of
the 2-dBKP system. One demonstrates how to derive the associated
Faber polynomials and their recursion relation via the Hirota
equations of 2-dBKP hierarchy.
Under the symmetry constraint \eqref{sym-con-1}, we solve
conserved densities by the derived Faber polynomials. Also, we
provide a set of hodograph equation of the dVN hierarchy, expanded
by the derivatives of its associated Faber polynomials.
Explicitly, we obtain the hodograph solutions to the dVN equation
as an example. \\
\indent For the more general symmetry constraint, we construct the
$2N$-component reduction system by the generalized Faber
polynomials and wrote down the corresponding hodograph equation.
However, the main difficulty is to find the explicit solutions of the
$2N$-reduction system (\ref{2N-hodo}).
We hope to address this problem elsewhere.

\subsection*{Acknowledgments}
The author J.H. Chang will thank Prof. Konopelchenko and Dr. Moro
for their stimulating discussions. This work is supported in part
by the National Science Council of Taiwan (Grant Nos. NSC
96-2115-M-606-001-MY2, J.-H. C., and NSC 96-2811-M-606-001, Y.-T.
C.).

\end{document}